\documentstyle[12pt,aasms]{article}
\newcommand{\beq}{\begin{equation}}
\newcommand{\eeq}{\end{equation}}
\newcommand{\etal}{{\sl et~al.~}}

\def \lta {\mathrel{\vcenter 
     {\hbox{$<$}\nointerlineskip\hbox{$\sim$}}}}
 
\begin{document}

\title{Self-Organized Large-Scale Coherence in Simulations of Galactic Star Formation}

\author{David Chappell and John Scalo}
\affil{Astronomy Department, University of Texas, Austin, TX 78712}
 
\begin{abstract}\baselineskip=12pt
It is often assumed that
galaxies cannot generate large-scale coherent star-forming activity without some organizing
agent, such as spiral density waves, bars, large-scale instabilities, or external perturbations
due to encounters with other galaxies.  We present simulations of a simple model of star
formation in which local spatial couplings lead to large-scale coherent, and even synchronized,
patterns of star formation without any explicit propagation or any separate organizing agent.  At
a given location, star formation is assumed to occur when the gas
velocity dispersion falls below a critical value dependent on the density.  Young stars inject
energy into the gas in their neighborhood, increasing the velocity
dispersion and inhibiting the instability.  A dissipation function continually ``cools" the
gas.  The stability of this local inhibitory feedback model is examined both analytically and
numerically.  A large number of two-dimensional simulations are used to examine the effect of
spatial couplings due to energy injection into neighboring regions.  We find that several
distinct types of behavior can be demarcated in a phase diagram whose parameter axes are the
density (assumed constant in most models) and spatial coupling strength.  These ``phases"
include, with decreasing density, a spatially homogeneous steady state, oscillatory ``islands,"
traveling waves of star formation or global synchronization, and scattered ``patches" of star
formation activity.  The coherence effects are explained in terms of the ability of the energy
injected  near a star formation site to introduce phase correlations in the subsequent cooling
curves of neighboring regions. It is
suggested that phases such as these, which depend mostly on the density, may occur in different
ranges of galactocentric distance within individual galaxies, and that galaxies as a whole may
evolve through different phases as the gas is gradually depleted by star formation, or because
the transient time to settle into a given phase may be very large.  In particular, the results
suggest that galaxies may develop large-scale or global oscillations or bursts in their star
formation rates during some stage of evolution, without the necessity for any organizing agent or
even propagation.  Such activity may explain the large range in present-to-past average star
formation rate ratios at a given morphological type found in two recent studies of disk galaxies,
and is consistent with the scattered low-level star formation seen in low surface brightness
galaxies and the outer disks of higher surface brightness galaxies.  The long ``incubation time"
($\sim10^9$ yr) required for synchronized global oscillations to develop suggests a possible
connection with observed redshift-dependent galactic phenomena.  The results are also 
applicable to the evolution of star formation in dwarf galaxies and in even smaller regions, such
as large molecular complexes.  However, true hydrodynamical spatial coupling, differential
rotation, and other effects  remain to be examined, and so the models must be interpreted as
merely suggestive of the type of self-organized pattern formation phenomena that might occur in
real galaxies.  The possible generic relation to synchronization phenomena in models for
excitable biological systems, especially systems of integrate-and-fire formal neurons,
is discussed.
\end{abstract}

\section{Introduction}
\baselineskip=24pt
Understanding the history of the global star formation rate (SFR) and its
spatial structure in galaxies is 
an important and open problem. Although the standard view of ``normal" galaxy evolution 
involves a SFR that decays monotonically in time (e.g. Searle, Sargent, and Bagnuolo 1973, Sandage
1986, Larson 1991 and references therein), there is considerable evidence suggesting
non-monotonic, irregular, or even intermittent SFR histories in several classes of galaxies. 
``Starburst" galaxies, which cover a large variety of morphologies, are certainly the most
prominent examples, and it is possible that all galaxies are starbursts at some time(s) in their
evolution.  While many of the most spectacular starburst galaxies are observed to be in
interacting systems, examples of isolated starbursts and non-nuclear starbursts are common,
especially at lower luminosities, leading some to speculate that such bursts could be due to
internal processes (e.g. Scalo 1987, Campos-Aguilar and Moles 1991, Coziol 1995).  
	
The evidence for nonmonotonic SFR histories in local (small redshift) ``normal" galaxies is 
more uncertain.  Observational constraints on the star formation history usually involve only the
ratio  of the recent to past average SFRs (commonly denoted as $b$) and 
provide no information on the 
detailed history $B(t)$.  For example, by comparing $H\alpha$ and 
broad-band observations, Kennicutt (1990, and references therein; also Kennicutt, Tambyn, and
Congdon 1994) found  that the ratio of the recent ($\lta10^7$yr) to past average SFR varied from 
approximately 0.02 to 7
for a sample of 210 spiral and irregular galaxies.  Much of this
variation is due to trends with galactic type in the sense that $b$
decreases toward early-type galaxies. This overall trend is 
frequently viewed as validation of the conventional picture of galactic
star formation in which $B(t)$ decays monotonically
in time with a faster decay rate for early-type galaxies.  
However, Kennicutt also finds significant variations of present-to-past average 
SFR within each
galaxy type, and the source of this spread is an unresolved question. 

Scalo (1987) pointed out that even  
this relatively small spread may be consistent with an irregular or even ``bursty'' global SFR.
Considering the uncertainties in the empirical SFRs, Kennicutt (1990) suggests that the
temporal SFR fluctuations are probably less than about a factor of two for intermediate-type
spiral galaxies, at least on timescales greater than several hundred million years, although they
are probably larger for dIrr galaxies.  If this is the case for smaller timescales, then the
variations could plausibly be interpreted as fluctuations in the number of small-scale ($<$kpc)
star formation sites within galaxies of a given morphological type, and there would be no
necessity to consider the possibility of globally coherent SFR variations.

However the situation is not so clear, partly because of the uncertain extinction corrections
that affect the H$\alpha$ equivalent width, and partly because of the relatively small sample
size at each morphological type.  Tomita, Tomita, and Saito (1996) investigated this question by
using the ratio of far infrared-to-blue fluxes as a measure of present-to-past average SFR for a
large sample of galaxies.  Tomita \etal found a large variation of this ratio for a given
morphological type, the spread being about an order of magnitude for Sb, Sbc, and Sc galaxies,
with the dispersion within each morphological type being significantly larger than the
differences in averages between morphological types.  A very similar result has been
recently found by Devereux and Hameed (1997).  Tomita et al. interpret their results as evidence
for strong global SFR fluctuations with a timescale less than about 10$^8$ yr.  

In addition,
there is evidence for large-scale SFR fluctuations in the Milky Way.  This evidence is based on
various estimates of
$B(t)$ from age distributions of stars presently in the  solor neighborhood, derived from, for
example, chromospheric ages, isochrone ages, or features in the  luminosity functions of main
sequence stars and white dwarfs (see Noh and Scalo 1990, Majewski 1993, sec. 3.2.1, and references
therein).  Since the samples include old stars which must have originated far from the solar
neighborhood (although there are no calculations to indicate just how far), these results suggest
large-scale SFR fluctuations in the Milky Way.  The timescales inferred for the fluctuations
are only upper limits, since all the methods involve smearing over times between 0.1 and 1
Gyr.  Similarly, the amplitudes are lower limits.

The existence of nonmonotonic SFR histories in at least some Local Group dwarfs is less
controversial, since color-magnitude diagrams for large numbers of individual stars can be used,
rather than the integrated light studies discussed above in connection with disk galaxies.  The
best-studied example is the Large Magellanic Cloud, in which star formation has apparently
erupted only intermittently, with most of its life spent in a relative lull (see Vallenari et al.
1996 and references therein).  For other Local Group dwarf galaxies, both irregular and
spheroidal, see van den Bergh (1994), Smecker-Hane et al. (1996), and references therein. 

Non-monotonic SFR behavior has also been inferred for galaxies at
intermediate redshifts.  Recent studies of the Butcher-Oemler effect (Butcher and Oemler 1978,
1984), in which the fraction of blue galaxies is larger on average for intermediate-redshift
clusters than for local clusters, indicate a progressive increase in the blue fraction out to at
least $z\sim1$ (Rakos and Schombert 1995).  The blue excess galaxies with strong emission lines
are probably experiencing elevated SFRs, while those with strong Balmer absorption lines can be
interpreted as ``post-starburst" systems (Dressler and Gunn 1983).  Close encounters between
cluster galaxies have been suggested as the cause of the elevated SFRs (e.g. Lavery, Pierce, and
McClure 1992, Couch et al. 1994).  However roughly half of the blue systems show no evidence for
interactions.  Dressler and Gunn (1983) suggested that bursts in these systems could be due to
motion through the dense intracluster medium.  Triggering of starbursts due to strong
intracluster medium shocks generated when subclusters pass through the main cluster has been
suggested by Caldwell and Rose (1997).  The physics behind these suggestions remains unclear. 
Kauffmann (1995a,b) has shown how the clustering history in bottom-up cold dark matter or perhaps
mixed dark matter cosmological models might account for the observations (see also Baugh, Cole,
and Frenk 1996).  However Schade et al. (1996) found that the surface brightness evolution of
field disk galaxies is indistinguishable from that of cluster galaxies, suggesting that it may be
incorrect to ascribe the redshift evolution to processes peculiar to rich clusters.  Butcher and
Oemler (1984) had already found that the effect occurs in open clusters as well as compact
clusters.
	A related phenomenon is the large population of faint blue galaxies found in deep galaxy
surveys (Kron 1982, Broadhurst, Ellis, and Shanks 1988, Tyson 1988).  These galaxies appear to be
dominated by late type/irregular systems (see Odewahn et al. 1996 and references therein; recent
studies of the Hubble Deep Field population can be found in Madau et al. 1996 and Mobasher et al.
1996).  Interpretations (especially concerning the question of the eventual fate of these objects)
include galaxy merging (Broadhurst, Ellis, and Shanks 1988, Kauffmann 1995a,b, Baugh et al. 1996;
but see Jones et al. 1997 for contrary evidence), evolution of the faint-end luminosity function
(Koo, Gronwall, and Bruzual 1993), a new population of dwarf galaxies (Cowie, Songaila, and Hu
1991), cycling SFRs, or continuous formation of new galaxies that fade after they reach a certain
age (Gronwall and Koo 1995, Pozzetti, Bruzual, and Zamorani 1996; see Bouwens and Silk 1996 for a
critique).  Conspicuous in all these speculations concerning Butcher-Oemler and faint blue
galaxies is the extremely vague or schematic nature of the star formation models invoked to
explain the observed behavior.  The same is true for local noninteracting starburst galaxies,
especially the non-nuclear starbursts, and one could make the same statement for galaxy
interaction simulations, since most assume an essentially ad hoc relation between SFR and
density, or between SFR and cloud collision frequency, etc., even though they are dynamically
quite sophisticated. 

The interpretation of these results in terms of global SFR fluctuations without a ``trigger"
poses a difficult theoretical problem:  How is it possible for star formation activity to become
coherent on large scales, when the duration of local events ($<10^8$ yr) is smaller than the time
for distantly-separated parts of a galaxy to communicate with each other?  The present paper
presents a possible answer, in the form of a simple model which is capable of self-organizing on
large scales.

The common interpretation of the observations in terms of a global SFR that only varies
monotonically and over large timescales has led most theoretical discussions to concentrate on
``one-zone" (no spatial degrees of freedom) self-regulating models.
 The view that galaxies are in near-equilibrium states with regard to their star
formation activity has been proposed by many authors.
 In particular, Franco and Cox (1983), Dopita and Ryder (1994),
and Wang and Silk (1994)
suggest that heating or ``stirring'' of interstellar gas due
to star formation provides feedback that might lead to a self-regulated
quasi-equilibrium state in which global properties change only over secular
time scales.  On the other hand, similar ``one-zone'' models (which also contain no
spatial information) that are allowed to explore nonequilibrium states often 
exhibit oscillations,
bursts, or chaotic behavior when time delays or sufficiently strong 
nonlinearities are
present (Ikeuchi and Tomita 1983, Scalo and Struck-Marcell 1987, 
Korchagin et al. 1988, Vazquez and Scalo 1991, Parravano \etal 1990, Parravano 1996). 
An important question is whether this nonequilibrium behavior is a 
purely local phenomenon or whether it can self-organize into coherent large-scale star
formation activity that fluctuates significantly in time.

In this paper we address this question by examining the global behavior
of a system of spatially coupled nonequilibrium one-zone models.
In section 2 we introduce and study the temporal behavior of 
a one-zone model in which massive stars and interstellar gas interact 
through two processes.  First, star formation is assumed to be
driven driven by the 
action of gravitational instabilities in which the star formation rate 
depends on the gravitational collapse time and 
the gas surface density (cf. Kennicutt 1989, 
Wang and Silk 1994).   These
instabilities arise as the velocity dispersion of the gas falls below
a critical value, which we take for illustrative purposes to be given by 
Toomre's Q condition (Toomre 1964);  thus the star formation
rate is a threshold function of the gas state variables.
Secondly, young stars inject energy back into the ISM, increasing the gas
velocity dispersion, which suppresses the instability and limits star formation.  
Thirdly, the gas
random motions driven by star formation decay at a rate given by a model dissipation 
function
appropriate for small-scale supersonic gas interactions.  Star formation feedback 
in this
model is inhibitory and could lead to a self-regulated state, although, as  
will be seen, the
self-regulation mechanism does not guarantee the existence of stable 
equilibrium solutions, and oscillations can develop.  

In section 3, we study the collective behavior of a system of these one-zone
models which are spatially coupled through stellar energy injection.  
We examine the manner in which
spatial degrees of freedom affect the behavior of model galaxies with
nonlocal inhibitory feedback of the star formation rate.
In particular, we wish to determine whether oscillations and bursts
of star formation
persist on a global scale when inhibitory coupling is present, and whether star formation can
self-organize to form coherent spatial structures.
A discussion of the results as they might apply to real galaxies is presented in section 4.

\section{Local Dynamics}
\subsection{One-Zone Model Equations}

We adopt a gravitational instability picture in which the star formation
rate $\dot{S}$ (with units of mass per unit area per unit time)
is set by the growth rate of the instability $\nu(\mu, c; t)$
(Wang and Silk 1994):
\beq
\dot{S}(t) = \epsilon \, \mu \, \nu(\mu, c;t)
						\label{eq:III_sfr}
\eeq
where $\epsilon$ is the star formation efficiency, $\mu$ is the gas surface
density, and $c$ is the gas velocity dispersion.  When the gas is stable,
star formation is shut off.  We adopt the stability 
condition of Toomre (1964), who showed
that gravitational instabilities in a rotating disk are suppressed
when $Q \equiv \kappa c / (\pi G \mu) > 1$, where $\kappa$ is the epicyclic
frequency.  
Observational evidence for large-scale star formation controlled by
the Toomre condition has been presented by Kennicutt (1989), although there
are known exceptions (e.g. Thornley and Wilson 1995).
Because we are interested in the coupling between local and global behavior,
we apply the Toomre condition locally, even though it strictly only applies
to larger scales.  We expect that any form of local threshold star
formation will give qualitatively similar results, and only adopt the
Toomre condition as an illustrative example.
The maximum growth rate for perturbations in a rotating sheet is given by
(Larson 1985)
\beq
\nu (\mu, c;t) = \left\{ \begin{array}{cl}
	\kappa \left[\left(\frac{\pi G \mu}{\kappa c}\right)^2 - 1 \right]^ {1/2} &
		\mbox{if $ c < \pi G \mu / \kappa$} \\
	0 & \mbox{otherwise.}
        \end{array}
\right.
						\label{eq:ode_growth_rate}
\eeq
We also considered cases in which the maximum growth rate for a spectrum is
truncated
at a given minimum wavenumber.  This would apply to geometries which possess a
maximum scale over which the instability can develop.  We found that
the form of our results do not depend on which growth rate is used.
For this discussion, however, the growth rate in 
eq. (2) will be assumed.

The energy equation governing the local (subgrid) gas velocity
dispersion is taken to be
\beq
\frac{d c^2}{dt} = - \alpha \, \mu (t) \, c^3(t) + 
\frac{1}{\mu(t)} 
	\int \dot{S}(t - t') \, g(t') \, {\rm d} t'
						\label{eq:ode_energy1}
\eeq
where the first and second terms on the right hand side correspond to
dissipation of the ``turbulent'' energy and energy injection by star formation,
respectively.  The basis for the adopted form of these terms is as follows.

We adopt a cloud-fluid picture to model the dissipation of the turbulent gas,
while recognizing that the representation of the real ISM as a system of
``clouds'' is problematical (Scalo 1990).  
In the absence of heating, the kinetic energy per unit mass associated with 
gas motions
decays as ${\rm d}c^2 / {\rm d}t \simeq - c^2 / \tau_{\rm diss}$, 
where $\tau_{\rm diss}$ is the dissipation time.
Assuming that the gas is distributed in 
clouds with radii $R_{\rm cl}$ and number density (number of clouds per
unit volume) $n_{\rm cl}$,
the characteristic dissipation time may be estimated as the collision time
$\tau_{\rm diss} = (n_{\rm cl} \pi R_{\rm cl}^2 c)^{-1}$. 
In terms of the mass surface density $\mu$, this becomes
$\tau_{\rm diss} = h m_{\rm cl} (\mu \pi R_{\rm cl}^2 c)^{-1}$ 
where $m_{\rm cl}$ is the cloud mass, $h$ is the galaxy scale height,
and we have used $\mu \approx n_{\rm cl} m_{\rm cl}h$. 
Thus, the cooling coefficient in eq. (3) is given by 
$\alpha \equiv \pi R_{\rm cl}^2 (m_{\rm cl}h)^{-1}$, assumed to be 
constant for simplicity, and the
dissipation time may be written as
\beq
\tau_{\rm diss} = \frac{1}{\alpha \mu c}.
						\label{eq:ode_tdiss}
\eeq
 
Massive stars inject kinetic energy into the surrounding ISM through
the action of winds and supernova explosions.  The rate at which this
energy is produced depends on the IMF and is, in general, a function of the 
time since 
the stars were born.  Denoting the rate of energy injection as a function of
the age of the star formation event by
$g(t)$, the total energy injected over the stars' lifetimes is
$g_{\rm tot} = \int g(t) {\rm d}t$.
Two characteristic times are considered.
The first, which
we designate by $\tau_{\rm d}$, represents a time delay between the onset of 
the gravitational
instability and the appearance of star formation and stellar heating due to winds.  
The second timescale is the duration of the stellar energy injection $\tau_{\rm w}$.  
We consider the two limiting cases:
$\tau_{\rm d} / \tau_{\rm w} >> 1$ and $\tau_{\rm d} / \tau_{\rm w} << 1$.
In the first case,
the kinetic energy is injected into the ISM on a much shorter time scale
than the gravitational collapse time and may be considered as occurring
instantaneously.  This limit could be applicable to individual supernovae which
inject energy on very short time scales.  In this case, the heating
function may be represented by 
\beq
g_{\rm sn}(t) = g_{\rm tot} \delta(t - \tau_{\rm d}).
						\label{eq:ode_SN_heat}
\eeq
In the second case, kinetic energy is produced at
a constant rate for a time $\tau_{\rm w}$ after star formation is initiated 
and
may be appropriate for stellar winds or the collective action of supernovae
from a cluster (see, for example, McCray and Kafatos 1987).  
In this case, it is assumed that the heating function takes the form 
\beq
g_{\rm w}(t) = \left\{ \begin{array}{cl}
	g_{\rm tot} / \tau_{\rm w} &
		\mbox{for $ 0 < t < \tau_{\rm w}$} \\
	0 & \mbox{otherwise.}
        \end{array}
\right.
						\label{eq:ode_wind_heat}
\eeq
We assume that the energy injected is proportional to the total mass
of young stars present.
 
Substituting the star formation rate (eq. (1))
into eq. (3), the energy equation becomes
\beq
\frac{\partial c^2}{\partial t} = - \alpha \mu c^3(t) + 
	\beta \int \nu(\mu, c; t - t') \, \tilde{g}(t') \, {\rm d} t',
			\label{eq:ode_energy}
\eeq
where we have defined $\beta \equiv \epsilon g_{\rm tot}$ and
$\tilde{g}(t) = g(t) / g_{\rm tot}$ to simplify the notation.
For the growth rate given by eq. (2) 
this equation has a single equilibrium solution when the velocity 
dispersion is below the star formation threshold, given by
\beq
\frac{\alpha \mu}{\beta \kappa} c_0^4 = 
\left[\left(\frac{\pi G \mu}{\kappa}\right)^2 - c_0^2\right]^{1/2},
\eeq
where $c_0$ is the equilibrium value of the gas velocity dispersion.
In the limit that $\pi G \mu/\kappa >> c_0$,
the solution for the equilibrium velocity dispersion is approximately
$c_0 \approx (\pi G \beta /\alpha)^{1/4}$, which is independent of both the 
gas surface density and star formation rate.  In this limit, the 
gravitational growth rate given in eq. (2) becomes 
$\nu \approx \pi G \mu / c_0 \propto \mu$. 
Thus eqs. (1) and (2)
each contribute one power of $\mu$ and 
a quadratic Schmidt law results with $\dot{S} \propto \mu^2$.
In the limit of small $\mu$, the equilibrium 
velocity dispersion approaches the star formation threshold and is given by 
$c_0 \approx \pi G \mu / \kappa$.

\subsection{Linear Stability Analysis}

Expanding the energy around the equilibrium solution as $c = c_0 + c'$,
and introducing
perturbations of the form $c' = \hat{c} e^{\omega t}$, results in the following
eigenvalue equation for the growth rate of the velocity dispersion perturbation:
\beq
\omega = - \frac{3 \alpha \mu}{2} c_{\rm 0} -
\frac{\beta}{2c} \left. \frac{\partial \nu}{\partial c} \right|_{c_0} \int 
{\rm e}^{-\omega t'} g(t') {\rm d} t'.
\eeq
As is typical for systems
with time delays, the eigenvalue equation is transcendental and a general
solution cannot be derived analytically.

Adopting the instantaneous heating function in eq. (5), the 
eigenvalue equation takes the simple (but still transcendental) form
\beq
\omega = - A - B {\rm e}^{- \omega \tau}
						\label{eq:ode_eigen_sn}
\eeq
where $A = 1.5 \tau_{\rm diss}^{-1} = 1.5 \alpha \mu c_0$ and 
$B = \beta / (2c) \partial \nu /\partial c$.  After substituting in the
growth rate, the B parameter becomes 
$B = 0.5 \kappa \beta [[\pi G \mu/ (c_0 \kappa)]^2 - 1]^{-1/2} 
(\pi G \mu / \kappa)^2 c_0^{-4}.$

In the limit of small $\tau$,
$\omega$ is real and negative and the system is stable against
small amplitude perturbations.  This is to be expected since the model equations
without time delays are stable.  Since the right-hand side of
eq. (10) is unconditionally negative, no positive real 
solutions exist;
however, complex solutions with real parts do exist.  To see this, we
decompose $\omega$ into real and imaginary parts
$\omega = \omega_r + i \omega_i$.  Substituting this into 
eq. (10), yields:
\begin{eqnarray}
\omega_r & = & -A - B {\rm e}^{-\omega_r \tau} {\rm cos} \; \omega_i \tau 
\nonumber \\
\omega_i & = & B {\rm e}^{-\omega_r \tau} {\rm sin} \; \omega_i \tau.
\end{eqnarray}
An instability occurs when the real part of $\omega$ crosses zero.
Setting $\omega_r = 0$ and solving for $\tau$, we find that the 
critical time delay at which small amplitude perturbations become
unstable is given by
\beq
\tau_s \equiv (B^2 - A^2)^{-1/2} {\rm cos}^{-1} \left( -\frac{A}{B} \right).
						\label{eq:III_bifurcate}
\eeq
This function is shown in Fig. 1a as a function of the gas 
surface density.
For $|A/B| \geq 1$ the equilibrium solution is stable against
small amplitude perturbations for all time delays.  Since this ratio may be 
rewritten as $A/B = 3 [ 1 - (\kappa c_0)^2/(\pi G \mu)^2]$, the condition
for unconditional stability becomes
\beq
Q_0 \equiv \frac{\kappa c_0}{\pi G \mu}\leq 2^{1/2},
\eeq
where $Q_0$ is the Toomre Q parameter evaluated at the equilibrium velocity
dispersion $c_0$.  To first approximation, $c_0 \approx \beta/\alpha$
near the point of unconditional stability, giving 
$Q_0 \approx (\kappa/\pi G \mu)(\beta/\alpha)$.
In the traditional application of the Toomre Q condition
to star formation in galaxies,
the gas velocity dispersion is assumed constant and independent of the
star formation rate itself (e.g. Kennecutt 1989).  Thus, given $c$, Q 
determines whether the gas is
unstable and if star formation proceeds.
According to the present model, however, the gas
velocity dispersion is coupled to the star formation rate, and, while the 
value of Q at a given time still controls the star formation rate, it may
vary in time.  Thus, the $Q_0$ parameter controls the 
stability of the star formation rate.
Stated differently, $Q_0$ dictates the stability of Q in time.

In the low density limit, $A/B << 1$, and the time delay at the bifurcation
point is 
$\tau_{\rm s} \approx B \approx 0.5 \beta^2 \alpha^{-1} \kappa^6 \mu^{-5}$.
(see Fig. 1c).

We also consider the constant heating function defined in 
eq. (6).  The eigenvalue equation becomes:
\beq
\omega = - A - \frac{B}{\omega} (1 - {\rm e}^{- \omega \tau})
\eeq
where $A$ and $B$ are defined as before.
Following the analysis in the above discussion, we find
that again there is a critical time delay above which perturbations
become unstable:
\beq
\tau_s \equiv (2B - A^2)^{-1/2} {\rm sin}^{-1} \left[ -\frac{A}{B} (2 B -A^2)^{1/2}\right].
						\label{eq:III_bifurcate2}
\eeq

\subsection{Discussion}

We find that the one-zone system undergoes a subcritical bifurcation
from a stable fixed point attractor to limit cycle oscillations as the 
time delay is increased above a critical value.  The solid lines in 
Fig. 2
represent the bounds on the energy oscillations as derived from numerical
integration using a sixth-order Runge-Kutta integrator.  Present in 
Fig. 2 are the characteristics of a
subcritical bifurcation:  the emergence of oscillations with large
amplitudes near the bifurcation point and a region of hysteresis in which 
both a fixed-point and a limit cycle attractor coexist, each with its own
basin of attraction.  The analytic bifurcation condition
(eq. (12)) applies only to small amplitude 
perturbations since it was derived from linear analysis.  
Its predicted values were confirmed through numerical integration.
The point $\tau_{\rm l}$
represents the bifurcation point for large amplitude perturbations and cannot
be investigated through the above linear analysis.

Fig. 1d shows the equilibrium star formation rate as a 
function of the gas surface density.  The flattening in the 
equilibrium star formation rate occurs at the point of unconditional
stability.  As discussed above, for large densities a quadratic Schmidt
law results.

In the low density limit, the turbulent dissipation timescale at the 
equilibrium energy scales as
$\tau_{\rm cool} \propto \alpha^{-1} \mu^{-2}$, since 
$c_0 \approx \mu/\kappa$ in this limit and 
$\tau_{\rm cool} \equiv (\alpha \mu c)^{-1}$.  Since in this limit
the heating rate depends only weakly on the density, the burst period
may be estimated roughly as the dissipation time, or $\tau_{\rm burst} \propto
\mu^{-2}$.  We find that the burst period of the simulations has a
somewhat steeper slope with
$\tau_{\rm burst} \propto \mu^{-2.6}$ (see Fig. 1c).
This discrepancy is due to variations in the heating rate as a function of
$\mu$, the form of which is not straightforward due to the presence of the
time delay in the star formation rate.

The adopted one-zone model is simple enough that it does not exhibit chaotic behavior for any
time delay, only limit cycles, a result that should be remembered in our examination of spatial
behavior below.

\section{Spatial coupling through stellar energy injection}

\subsection{Model Equations}

We now investigate a two-dimensional spatial system in which the above
one-zone models are coupled through the stellar heating function, whose effect is to increase
the gas velocity dispersion. No mass transport is allowed.  Thus, the cooling function and
star formation law remain unchanged; only the stellar heating function
must be modified to include the spatial dependence.

We model the spatial distribution of the heating energy (which defines the 
coupling between the one-zone models) by the function $f(|{\vec x}|)$.
We consider a nearest-neighbor coupling function normalized such that
$\int f(| \vec{x} |) {\rm d} \vec{x} = 1$.
Including the spatial heating function and assuming instantaneous heating
$\tilde{g}(t) = \delta(t - \tau_d)$, the energy equation becomes
\begin{eqnarray}
\frac{\partial c^2(\vec{x}, t)}{\partial t} & = & -\alpha \mu(\vec{x}, t) 
c^3(\vec{x}, t) +  
\frac{\epsilon g_{\rm tot}(1 - \eta)}{\mu(\vec{x}, t)} 
\dot{S}(\vec{y}, t - \tau) + \nonumber \\
	&  &  \frac{\epsilon g_{\rm tot} \eta}{\mu(\vec{x}, t)} 
\int \int \dot{S}(\vec{x}^\prime, t - \tau) f(| \vec{x} - \vec{x}^\prime |) 
{\rm d} \vec{x}^\prime,
						\label{eq:III_energy_2D}
\end{eqnarray}
where $g_{\rm tot}$ represents the total energy injection rate, $\eta$
represents the fraction of the heating which is non-local, and $\epsilon$
is the star formation efficiency.
The second term on the right hand side of this equation represents the
local heating and the third term non-local heating.

\subsection{Linear stability analysis}

Assuming a constant homogeneous gas distribution 
and substituting the star formation rate into
the energy equation, the following  
equation results:
\beq
\frac{\partial c^2(\vec{x}, t)}{\partial t} = -\alpha \mu
c^3(\vec{x}, t) +  (1 - \eta) \beta 
\nu(\vec{x}, t - \tau) + \eta \beta
\int \int \nu(\vec{x}^\prime, t - \tau) f(| \vec{x} - \vec{x}^\prime |) 
{\rm d} \vec{x}^\prime,
\eeq
where the substitution $\beta \equiv \epsilon g_{\rm tot}$ has been made.
Because of the normalization of the heating function, the equilibrium 
solution for the spatial system is the same as for the one-zone model.

Linearizing this equation and introducing perturbations of the form
${\rm e}^{\omega t + i k x}$, we find that the dispersion relation takes
the same form as the eigenvalue eq. (9):
\beq
\omega = - \frac{3 \alpha \mu}{2} c_{\rm 0} -
\frac{\epsilon \beta}{2 c} \left. \frac{\partial \nu}{\partial c} \right|_{c_0} 
\hat{F}(k) {\rm e}^{-\omega \tau}, 
\eeq
where 
$\hat{F}(k) = \int \int f((x^2+y^2)^{1/2}){\rm e}^{i k x}{\rm d}x{\rm d}y$ 
is the one-dimensional Fourier transform of the 
heating kernel $f(|\vec{x}|)$.  The stability of the system depends
on the effective heating radius through the Fourier transform of the
heating kernel.

\subsection{Simulation results}

Equation (16) was integrated using a 6th order
Runge-Kutta integrator on a Sun workstation.
Simulations were run on 
$64^2$ and $128^2$ lattices with a lattice spacing 
of 78 pc, giving simulated areas of $5^2$ kpc$^2$ and $10^2$ kpc$^2$, respectively.
The time step, time unit and time delay are set at $10^6$, $10^7$ and
$1.5 \times 10^7$ years respectively.  The time delay was chosen to correspond to the time from
which the gravitational instability begins to the time at which the stellar energy injection is
transferred to the surrounding gas.  
A Gaussian coupling function was  adopted with a standard
deviation of
$1.5
\Delta x$.  Thus, the effective diameter of the heating kernel is 120 pc.  The physical scale
chosen is somewhat arbitrary; we adopt these values mostly to put
the problem in a galactic-scale context.  For example, the simulations could
be scaled down to represent the stirring of the interiors of GMCs by 
low-mass protostellar winds.  For the galactic-scale context, one measure of the
scale of influence of massive stars on their environment is the size of
wind-blown or supernova shells, which range from a few parsecs to as much
as a kiloparsec but with an average from 100 to 200 pc (Oey and Clarke 1996).
The effect of varying the effective size of the heating kernel (see eq. 18) is discussed in sec.
3.5.  We set the value of the cooling coefficient by adopting typical cloud properties (see the
discussion preceeding eq. (4)). Assuming the cloud internal density,
cloud radius, and gas scale height in the galaxy are 
$n_{\rm int} = 30$ cm$^{-3}$, $R_{\rm cl} = 5$ pc, and h = 100 pc respectively,
the cooling coefficient in simulation units becomes $\alpha = 0.3$.
The initial velocity dispersions were chosen such that the phases of the
initial oscillations (or decays in the stable cases) were uncorrelated across the grid, so that
any spatial coherence that arises during evolution cannot be ascribed to the initial
conditions.

All simulations were run for approximately $5\times10^9$ yr, in order to get past any
long-term transients (which are known to be a common phenomenon in coupled lattice models for
complex systems) and to search for non-transient behavior that may occur on a large timescale. 
The long integration times turned out to be essential for revealing the onset of
global synchronization discussed below.  Such long integration times would not be feasible for
high-resolution hydrodynamic simulations. 

Table 1 shows the values of the parameters which were varied and a few statistics
for 38 model runs.   Figure 4 shows a ``phase diagram," the axes of which are the parameters
coupling strength and density $\mu$, for the models, and five regimes which exhibit qualitatively
different behavior. Figure 5 shows snapshots of the spatial structure of star formation
activity for eight points in the phase diagram of Fig. 4.  (For an illustrative time
sequence of structures, see Fig. 8, which is discussed in sec. 3.5. below.)  Figure 6 shows the
global evolution of the star formation rate and the energy for representative models in each of
the phases.

For $\mu/\mu_0 > 1.17$, a spatially homogeneous steady state is reached 
(labeled as phase V in Table 1),
analogous to the stable equilibrium solution found for the one-zone model.  All the sites in the
simulation attain equal, and constant, values of the SFR and gas velocity dispersion.  The
critical value of
$\mu/\mu_0$ is independent of the strength of the coupling coefficient $\eta$ since the
equilibrium solution is stable for each one-zone model.

As $\mu$ decreases below the bifurcation point at 1.17, islands of limit-cycle 
attractors appear at the locations where the initial trajectories fall 
in the limit-cycle
attractor basin.  These oscillatory islands are stationary and are surrounded by a 
smooth sea of steady state behavior (phase IV, see
G and H in Fig. 5 and Table 1).  As the density is further 
decreased the
limit-cycle attractor basin grows and along with it the spatial filling
factor of unstable islands.  For $\mu /\mu_0 < 1.1$ stable fixed-point
solutions no longer exist.  All the lattice sites are unstable with respect to local
oscillations, and the lattice develops globally synchronized oscillations or traveling waves of
star formation, depending on the values of both
$\mu$ and the coupling coefficient $\eta$.  Global synchronization occurs
at large values of the coupling coefficient.  This behavior is only established after a long
transient time (see Table 1).  The degree of synchronization is measured by the ratio
$\chi_{\rm SFR} \equiv <\sigma_{\overline{\rm SFR}}>  /
\sigma_{\overline{\rm SFR}}$, where $\sigma_{\overline{\rm SFR}}$
is the standard deviation of the fluctuations of the global star formation
rate in time and
$<\sigma_{\rm SFR}> $ is the temporal average of the standard 
deviation of the spatial fluctuations.  Thus, small values of
$\chi_{\rm SFR}$ indicate that the spatial fluctuations are, on average,
smaller than the global variations, signaling that the oscillations
are correlated.
We choose $\chi_{\rm SFR} < 0.25$ as the criterion to identify
global synchronization. The precise value is not crucial, since we are
simply interested in identifying whether the global dynamics develop
significant oscillations.  Animated visualizations of the simulations verified that the adopted
criterion does correctly identify the occurrence of global synchronization.

The large-scale coherence in the synchronized phase develops as follows.  A site that cools
below the star formation threshold injects energy additively to its neighboring sites.  Without
the added energy, neighboring sites with energies above the threshold would cool without
correlation, reaching the threshold at times that are uncorrelated.  The addition of the same
(or similar) energy to these neighboring sites pushes them farther from the threshold, and the
nature of the dissipation function, which gives $c^2\sim t^{-2}$ in the absence of heating,
implies that they will arrive at the threshold at times that are less separated than would be
the case without the energy input.  In other words, the energy injection introduces some
correlation into the initial conditions for their cooling curves.  These neighboring sites reach
the star formation threshold at different times, but the star formation activity is now
correlated in time.  As this process repeats over several generations, the correlations become
stronger, leading to synchronization.  After a sufficient time, the system evolves to a state in
which local patches of correlated star formation induce correlated heating, and then cooling, in
the surrounding areas.  So even though propagation is not explicitly built into the model, the
induced correlations lead to star formation that in effect propagates in rings which subsequently
heat themselves and larger surrounding rings.  When the dissipation time ($\tau_d\sim1/\mu c$, eq.
(4)) is large (i.e. the density small enough), this effect can push all the sites on the grid
above the threshold, and then the synchronizing effect of the cooling curve can synchronize the
entire grid.  Apparently this can only occur if the spatial coupling coefficient $\epsilon$ is
larger than some critical value.  Otherwise the system settles into a configuration dominated by
traveling waves of star formation.  The development of synchronization is gradual, requiring
($0.7-3)\times10^9$ yr for the models examined here (see Table 1).

As the density is decreased further, the local dynamics become increasingly 
bursty and the local star formation duty cycle accordingly decreases.  Since heating
due to star formation provides the coupling, the decrease in the local star
formation duty cycle with decreasing gas density reduces the amount of time
that neighboring regions have to communicate.  Thus, it becomes more 
difficult for
the system to synchronize at lower densities and the
spatial pattern of star forming sites appears patchy and uncorrelated
(phase I).  The global star formation rate then settles into an
equilibrium state with only small amplitude oscillations 
(see Fig. 6).

\subsection{Effects of Density Fluctuations}
All the simulations described so far take place on a constant-density background, so the only
``noise" present is due to the fluctuations in velocity dispersion which arise both because of
the random initial conditions and the subsequent spatial variations in the rates of the star
formation, neighborhood heating, and cooling.  We examined the effect of introducing spatial
fluctuations in the density, which is equivalent to including a spatially stochastic component
to the threshold criterion for star formation and the cooling function.  The density
fluctuations were noise drawn from a uniform probability distribution with specified standard
deviation.  For this preliminary investigation, the spatial distribution of density fluctuations
is fixed throughout the simulation.  We find that the inclusion of relatively small amplitude
density fluctuations does not qualitatively affect the above results.  However, as the amplitude
of the fluctuations in increased, the spatial structures become smeared and ``mottled" in
appearance, although they retain similar spatial scales of overall coherence as in the cases
without density fluctuations.  

This behavior is illustrated in Fig. 7, where the smooth star formation
distribution and the traveling star formation waves seen in phases III and II, respectively,
become scattered and disorganized when large density fluctuations are introduced.  This
small-scale decoherence is accompanied by a decrease in the amplitude of the global oscillations,
since nearby local regions are now out of phase and may have very different star formation rates,
so any large-scale spatial coherence in the structure is largely masked by averaging over these
small-scale fluctuations.  Thus, in the presence of sufficiently large density fluctuations the
global star formation rate becomes more nearly constant in time, independent of the values of
the coupling coefficient $\epsilon$ and the average gas density.  This behavior results from the
dependence of the oscillation period of the one-zone models on the local gas density: 
With the inclusion of the density fluctuations the model becomes equivalent to a system of
coupled oscillators with a probability distribution of natural frequences.  Of course in a more
realistic model the density fluctuations would be allowed to evolve in time, in a manner which
must to some degree be coupled to the velocity dispersion (and to the overall flow velocity,
which is omitted from this non-hydrodynamic model).  It is not clear how to model such a
coupling, and it is possible that in such a model the effect of density fluctuations might be to
enhance or decrease the coherence, depending on the nature of the coupling.  However, we suspect
that the example chosen here is the ``worst case scenario" for disrupting the synchronization.

\subsection{Effects of Varying the Energy Injection Radius}
It was shown in sec. 3.2 above that the linear stability of the spatially coupled system depends
on the size of the region in which energy is deposited, which we refer to as the ``heating" or
``stirring" radius.  In the standard models discussed above, the non-local heating function was
taken to be a Gaussian with a small standard deviation of 1.5 $\Delta$x ($\Delta$x = grid
spacing).  We have examined a few models with larger heating radii, with standard deviations up
to 4 $\Delta$x, and find that this change significantly enhances the ability of the system to
synchronize.

Figure 8 shows the time evolution of two models to the left of the synchronization boundary of
Fig. 5.  The top row of each pair shows the standard models with heating radius 1.5 $\Delta$x
while the lower panels show the behavior for a heating radius of 4 $\Delta$x.  The top pair of
panels are for parameters of the model labeled A in Fig. 5, just to the left of the
synchronization boundary.  It can be seen that the increase in heating radius causes this
transition case, which already displayed coherent but not globally synchronous behavior, to
become coherent on very large scales, and it in fact becomes essentially synchronized.  The
mechanism of coherence involving nearly circular regions of propagating inhibition,
discussed in sec. 3.3, can be seen.  The lower two panels correspond to the point to the left of
point A in the phase diagram.  The upper panel of this pair shows the star formation activity at
four times for the 1.5 $\Delta$x standard model, which gives scattered star formation; the
lower panel shows the 4 $\Delta$x case.  The transformation into large-scale coherence is
dramatic, and illustrates the sensitivty of the size of the synchronization region in the phase
diagram to the heating radius.  Notice that the heating radius is still a small fraction of the
size of the system; this fraction has only been increased from 0.024 to 0.062.

We also re-ran the model corresponding to point C in the phase diagram using the larger heating
radius.  The standard model resulted in traveling waves of star formation.  With the larger
heating radius the behavior was qualitatively the same, except that the width and separation of
the fronts was magnified.

\section{Discussion}

The spatially coupled system of threshold one-zone models is found to 
exhibit synchronized oscillations, quasi-equilibrium behavior, and other ``phases" that depend
on the parameters of our model problem.  Global oscillations develop when the gas surface density
lies in an intermediate range and the  spatial coupling is sufficiently strong.  The high density
limit for the emergence of global oscillations is controlled by 
$Q_0 \approx \kappa \beta / (\pi G \mu \alpha)$ where $\beta/\alpha$ is
the ratio of the heating to cooling rates.  This parameter is closely
related to the Toomre Q parameter.  However, since we have adopted the Toomre condition only as
illustrative of threshold star formation, we do not place much significance on the actual values
of the critical densities, but expect similar qualitative behavior for other threshold
prescriptions.  The main point is that such models can display self-organized spatially coherent
activity without the need for any separate organizing agent (like a galaxy encounter or a bar)
or even explicit propagation of star formation.

Because we have claimed that our model involves no ``explicit" propagation of star formation, 
while any transport model (even pure diffusion or advection) can be said to involve
``propagation" in some sense, it is important to clarify the relation of our model to the studies
of propagating star formation models by Gerola and Seiden (GS; see Seiden and Gerola 1982 for a
review).  First of all, our model does not employ any stochastic formulation for the occurence of
either ``spontaneous" or ``propagating" star formation.  Our model is completely deterministic and
star formation only occurs when the gas velocity dispersion, responding to heating by stars and
the adopted cooling function, falls below a certain threshold.  While the present model does not
include any explicit propagation of star formation, the heating of nearby regions and their
subsequent cooling can lead to an effective propagation of inhibition followed by star formation,
and so the models are related in that sense.  However the spatial connectivity is purely
inhibitory here, with star formation occurring only when cooling can overcome the heating due to
previous star formation.  This heating of the neighborhoods of  star formation sites effectively
leads to a distribution of ``refractory times" (times during which stars cannot form) that could
be thought of as analogous to the (single-valued) refractory time used in the GS model, but the
GS refractory time was applied to a site that had just formed stars, not its neighborhood.  The
present model is therefore not symmetric with the GS model, in the sense in which Freedman and
Madore (1984) discussed the similar behavior of propagating inhibitory and excitatory models for
spiral structure.  Another point is that the bursts found in the GS simulations for dwarf
galaxies of small sizes were not due to any synchronization, but occurred because the time to
fill up a small galaxy with propagating star formation could be smaller than the adopted
refractory time.  In our model there is a similar effect once a model becomes synchronized, but
it requires the gradual process of phase-locking, discussed earlier, before it occurs; and when
it does occur, the coherent region can occupy a much larger area than the GS model bursts, at
least for the respective adopted values of the parameters.  The prediction of a strongly
increasing burstiness duty cycle with decreasing galaxy size in the GS model leads to the
prediction of huge numbers of very small but very low surface-brightness galaxies (Tyson and
Scalo 1988), in excess of observed numbers based on several HI surveys.  The present model makes
no such prediction about galaxy size dependence, which is a subject we have not yet
investigated.  The focus of our calculations is simply to demonstrate the possibility of
self-synchronized behavior with a long onset time.  In that sense the long-term oscillations
found in some spiral galaxy GS-type models by Gerola, Seiden, and Schulman (1980) is a behavior
more akin to the phenomena discussed here; to our knowledge, an explanation of this oscillatory
behavior has not appeared, and deserves further study.

Our result that global oscillations can develop is in contrast
to the coupled oscillator system studied by Parravano et al. (1990) and Parravano (1994).  
They considered
a one-dimensional array of self-regulating one-zone models, which are
coupled through heating due to radiation in the 912 - 1100 \AA $\;$ band.
In this model the radiation 
controls the rate of evaporation and condensation of small clouds, which
determines the star formation rate.
They found that even when only two one-zone models are coupled, the
individual systems oscillate out of phase, leading to nearly constant
``global'' averages.  Parravano (1994) investigated a coupled 50-oscillator array and again
found out-of-phase behavior, although his Fig. 1 shows that some long-range correlations do
develop.  On the other hand, we find robust global oscillations in a large fraction of parameter
space even with  essentially nearest-neighbor coupling.  The difference apparently lies in our
adoption of a threshold condition for star formation and the effectiveness of the cooling
function in introducing correlations between neighboring regions to which energy has been added
by star formation.

We find that when global synchronization does $not$ occur, two spatial
organizations of the star formation activity are possible.  When the
gas surface density is small, star formation is locally very bursty
and patchy.  In this case, the nearly-constant global behavior is
the result of many spatially and temporally uncorrelated bursts.
At larger gas surface densities, wave trains of star formation develop.  
When the corresponding wavelength is considerably less that the integral
scale of the system, the global averages are again approximately
constant in time.  At still larger densities there exists a density range in which oscillatory
``islands" or clusters of activity, surrounded by a sea of equilibrium behavior, occur.  It is
likely that these clusters become synchronized by the same heating/cooling--induced
phase-looking mechanism discussed in connection with global synchronization.

Since the average gas surface density in disk galaxies typically decreases 
faster than the epicyclic frequency with increasing galactocentric radius
(e.g. Kennicutt 1989), 
this model suggests that a transition from homogeneous star formation, to small-scale 
synchronized clusters, to star formation waves or global synchronized oscillations, 
to patchy star
formation, may occur with increasing galactocentric radius.  In particular,
we suggest that this trend may explain the transition from organized
star formation in the inner regions of galaxies to the scattered patches
of star formation in the outer regions (e.g. Ferguson \etal 1993).
While variations in average $\mu$ from galaxy to galaxy are typically
less than the variations within a given galaxy, this progression of
behaviors could also represent an evolutionary sequence as the average
gas density decreases due to gas depletion by star formation.
Although this picture is consistent with the traditional
view that the overall tendency of the star formation rate in disk galaxies 
is to decrease with time, it predicts that there could be a stage during
the beginning or middle of a galaxy's life in which global oscillations develop.  Self-organized
global SFR oscillations could, if the amplitudes are sufficient, contribute to the large spread
in present-to-past average SFR ratios at a given morphological type found by Tomita et al. (1996)
and Devereaux and Hameed (1997). In addition, a disk galaxy's average gas column density may
depend on environmental effects or initial conditions, as perhaps has occurred in low-surface
brightness (LSB) galaxies.  If the column densities of LSB galaxies are sufficiently small, the
present models predict that star formation will only occur at scattered sites, with a small
global rate, in rough agreement with what is observed (See Bothun, Impey, and McGaugh 1997 for a
review and references).  Thus, these models predict that the {\em large-scale} spatial
and temporal behavior of the galactic SFR (constant, oscillatory, or bursty) should depend
primarily on the gas surface density.

A particularly interesting feature of the present simulations is that it takes a long time
($0.7-3\times10^9$ yr for the models examined here) for the phase correlations to grow
sufficiently for global synchronization to occur.  This suggests that galaxies born at a density
conducive to synchronized behavior may not develop SFR oscillations until ages corresponding to
intermediate redshifts.  A connection with the Butcher-Oemler effect and related
redshift-dependent phenoma discussed in sec. I is possible.\footnote[1]{A relatively long
``incubation period" before a burst had been proposed by Vazquez and Scalo (1989), but the
one-zone models that gave that behavior were very different from the spatially-coupled models
investigated here.  In the Vazquez and Scalo models, infalling gas clouds were supposed to stir
up the disk sufficiently that clouds were ``shredded" during collisions, and the ``incubation
time" reflected the long time required to rebuild a population of clouds massive enough to form
stars.  The question of spatial coherence or synchronization was not addressed.}

While this model has provided one example of how the non-equilibrium
behavior of a one-zone model can be affected by the introduction of
spatial couplings, it has several deficiencies.  
First, although we examined the effect of fixed density fluctuations, we do not know how the
density field should be coupled with the velocity dispersion field.  Depending on the coupling,
a density field with spatial fluctuations could reduce or enhance the ability of the system to
synchronize.  Second, the model ignores differential galactic rotation, which could significantly
alter the results by introducing anisotropy.  Third, we have not considered stochastic effects
(e.g. a probability distribution of inhibition radii or other quantities), which might act to
de-correlate the star formation activity. In addition, the coupled system still lacks
hydrodynamic couplings. (Hydrodynamic simulations with threshold star formation and
wind-driven shells will be presented elsewhere; Chappell and Scalo 1997, in preparation.)  Thus,
effects of gas motions, instabilities, turbulence, etc. on the predictions of the original
one-zone model could not be investigated.  Given these caveats, and the huge distance between
{\it any} models and real galaxies, we still think the results may have observational implications
since they demonstrate, contrary to some expectations, that it is possible for models of locally
inhibitory threshold star formation events to  ``self-organize'' into large-scale coherent
structures, and that the development of this behavior may take a long time, suggesting a
connection with observed redshift-dependent evolutionary phenomena.

Finally, we note the generic similarity of our results to models for biological phenomena that
involve collective synchronization of local oscillations (e.g. Kuramoto 1991, Chawanya et al.
1993). A large number of references can be found in the recent paper by Tanaka, Lichtenberg, and
Oishi (1997).  As a specific example, van Vreeswijk (1996) has shown that populations of
nonlinear oscillators with inhibitory couplings (as in the present models) can lead to the
development of a number of synchronized clusters, reminiscent of the ``phase IV" behavior found
here. 

Particularly intriguing is the fact that our model resembles a network of coupled ``integrate
and fire" threshold units often used to investigate temporal synchronization in pacemaker cells,
the cortex, and other biological systems (e.g. Mirollo and Strogatz 1990, Vanvreeswijk and Abbott
1993, Arenas and Vicente 1994, Herz 1995, Park and Choi 1995). In the neural network case the
neuron ``fires" (produces an action potential) only if its membrane potential is greater  than a
certain threshold.  In typical models the potential of a particular cell is increased by the sum
over outputs of neighboring cells, but decreases (exponentially) in the absence of connectivity
due to its ``leakage" current.  It is the signals from the neighborhood that push the cell to its
firing threshold (i.e. the coupling is excitatory), while the cell moves away from the threshold
due to its own leakage.  In the present SF model, firing (star formation) occurs only when the
velocity dispersion falls below a threshold.  Cooling is analogous to leakage in the neural case,
since it is monotonic and does not depend on couplings to other cells (it is purely local), but
the cooling drives a cell to fire, while leakage in the neural case drives a cell away from the
firing threshold.  The heating of a cell due to neighboring active cells is similar to the
integration over neighboring cell outputs in the neural case, but the effect on firing is again
reversed: the integrated field is inhibitory, not excitatory.  The ``cool-and-fire" prescription
leads to a range of refractory times.  In these terms the star formation model is a kind of
mirror image of the integrate-and-fire neural model.  It is therefore quite interesting that
systems of integrate-and-fire networks can display synchronized behavior.  In particular, with
finite-range couplings, some networks evolve to phase-locked clusters of synchronized neurons
(see Herz 1995 for references).  This result is usually interpreted as related to the idea that
synchronized cortical neurons ``bind" stimulus features together.  If the synchronization in the
star formation case could be shown to be associated with a Lyapunov function, as has been shown
for integrate-and-fire neural network models, this would imply the possibility of a statistical
mechanical formulation of models for star formation.  More generally, and speculatively,  
the formal resemblance of the models,
the common finding of a transition in a phase diagram between synchronized and asynchronous
behavior, and the development of global synchronization through the growth of synchronized
clusters, suggests the possibility of a generic connection between the spatio-temporal behavior
of star formation in galaxies and a variety of biological phenomena, including information
processing.

This work was supported by NASA grant NAG 5-3107.

\eject
Figure 1: Properties of the one-zone model for $\kappa=1,\ \alpha=0.3$\quad a) Critical time
delay above which small amplitude perturbations lead to bifurcation (see eq. (12) in the text). 
For
$\mu>1.1$ the equilibrium state is stable for any time delay.  As $\mu$ decreases the
characteristic turbulent dissipation time becomes long compared to the characteristic time for
stellar heating and the system becomes unstable for smaller time delays.  b) Bounds on the
energy oscillations as a function of density for $\tau=1.23$ and $\alpha=0.37$.  Above $\mu=1$
the equilibrium solution is stable and the figure shows the dependence of the equilibrium energy
as a function of $\mu$.  c) Oscillation period based on numerical integration for a time delay
$\tau=1.5$.  For small densities the oscillations grow increasingly bursty and the period
increases.  d) Equilibrium star formation rate.  For densities above 1.1, a quadratic Schmidt
law results.  The change in the power-law slope occurs at the point where the dynamics become
unconditionally stable.

Figure 2:  Subcritical bifurcation from a stable fixed point attractor to limit cycle
oscillations.  The solid lines represent the bounds on the energy oscillations derived from
numerical integration.  $\mu/\kappa=1,\ \alpha=0.3$.  For $\tau_1<\tau<\tau_s$ both the
fixed-point and limit cycle attractors are stable, each with its own basin of attraction.

Figure 3:  Star formation rate (dotted lines) and gas internal energy (solid lines) for the
one-zone model as a function of time.  $\kappa=1,\ \alpha=0.5,\ \tau=1.5$.  Just below the
bifurcation point, the oscillations are symmetrical and the star formation rate remains above
zero most of the time.  As the density is decreased, star formation becomes bursty and the
energy oscillations become asymmetrical.

Figure 4:  Phase diagram for the spatially coupled system.  $\kappa=1,\ \alpha=0.5,\ \tau=1.5$. 
The models investigated are indicated by x's.  The solid and dotted vertical lines correspond to
the bifurcation point for small and large amplitude perturbations.  Between them lies a region in
which both fixed-point and limit-cycle attractors can coexist.

Figure 5:  Star formation activity at time $t=500$ for eight points in the phase diagram in
Fig. 4.  White and black represent the highest and lowest star formation rates in a given
image; the images are not on a common intensity scale.  The vertical pairs show two examples for
each of the four dynamical phases.  The labels A--H refer to the corresponding labels in the
phase diagram and in Table 1.  The values of the gas surface densities and coupling coefficients
are given in Table 1.

Figure 6:  Global star formation rate (dotted lines) and gas internal energy (solid lines) as a
function of time for models A, C, E, and G which are representative of phases I, II, III, and
IV, respectively.  $\kappa=1,\ \alpha=0.3,\ \tau_d=1.5$.  The parameter values for each model
are given in Table 1.  Examples of the spatial star formation patterns for each of these
models are shown in Fig. 5.

Figure 7:  Snapshots of the instantaneous star formation activity in models which would be in
the synchronized oscillation phase III (top row) or traveling wave phase II (bottom row) in the
absence of density fluctuations (leftmost images).  The fixed density fluctuation standard
deviation increases from left to right, with values of 0, 0.05, 0.1, and 0.4.

Figure 8:  Time evolution (left to right) of models showing the effect of increasing the
standard deviation of the gaussian nonlocal energy injection function from 1.5$\Delta$x (top row
of each pair of panels) to 4$\Delta$x (bottom row), where $\Delta$x is the lattice spacing. 
The top pair of panels are for parameters of model A in Fig. 5, just to the left of the
synchronization boundary, while the lower pair of panels correspond to the point to the left of
point A in the phase diagram.  In both cases the increase in stirring radius enhances the
coherence and induces synchronization.

\end{document}


\begin{deluxetable}{lrr|rrrrrrrrrrrrr}
\tablecolumns{16}
\tablewidth{0pc}
\tablecaption{Model Runs}
\tablehead{
\colhead{} & \multicolumn{2}{c}{Parameters} &   \colhead{}   & 
\multicolumn{12}{c}{Statistics} \\
\colhead{} & \colhead{$\eta$} & \colhead{$\mu$}   & \colhead{$\overline{E}$} &
\colhead{$\chi_{\rm E}$} & \colhead{$\overline{SFR}$ }    & 
\colhead{$\sigma_{\overline{\rm SFR}}$}   & \colhead{$<\sigma_{\rm SFR}>$} & 
\colhead{$\chi_{\rm SFR}$}  & \colhead{$\overline{ff}$}  & 
\colhead{$ff_{\rm min}$} & \colhead{$ff_{\rm max}$}  & \colhead{$N_{\rm G}$} & 
\colhead{$\overline{T}$}  & \colhead{$\tau_{\rm sync}$} & \colhead{Phase}\\ 
\colhead{}  & \colhead{(1)} & \colhead{(2)} & \colhead{(3)} & \colhead{(4)} & 
\colhead{(5)} & \colhead{(6)} & \colhead{(7)} & \colhead{(8)} & \colhead{(9)} & 
\colhead{(10)} & \colhead{(11)} & \colhead{(12)} & \colhead{(13)} 
& \colhead{(14)} & \colhead{(15)}
}
\startdata
  & 0.8 & 2.00 & 1.43 & 0.06 & 2.675 & 0.542 & 0.089 & 0.16 & 0.98 & 0.08 & 1.00 & 4.2 & 4.8 & -- & V\\
E & 0.8 & 1.00 & 0.90 & 0.08 & 0.314 & 0.272 & 0.019 & 0.07 & 0.64 & 0.00 & 1.00 & 0.6 & 4.9 & 71 & III \\
F & 0.8 & 0.70 & 0.57 & 0.23 & 0.081 & 0.113 & 0.025 & 0.22 & 0.34 & 0.00 & 1.00 & 0.4 & 6.6 & 274 & III \\
  & 0.8 & 0.68& 0.54 & 0.36 & 0.075 & 0.097 & 0.040 & 0.41 & 0.34 & 0.00 & 1.00 & 0.6 & 6.9 & -- & II\\
  & 0.8 & 0.67& 0.52 & 0.71 & 0.072 & 0.075 & 0.064 & 0.86 & 0.34 & 0.00 & 0.90 & 1.3 & 7.0 & -- & II\\
  & 0.8 & 0.65& 0.49 & 5.75 & 0.053 & 0.012 & 0.088 & 6.96 & 0.29 & 0.20 & 0.40 & 20.1&  7.3&  -- & II\\
  & 0.8 & 0.60 & 0.40 &21.86 & 0.033 & 0.002 & 0.063 &26.83 & 0.25 & 0.21 & 0.29 & 98.4&  6.3&  -- & I\\
  & 0.8 & 0.50 & 0.28 &22.72 & 0.014 & 0.001 & 0.036 &35.85 & 0.15 & 0.12 & 0.17 & 137.6&  5.1&  -- & I \\ \hline
  & 0.6 & 1.00 & 0.91 & 0.22 & 0.306 & 0.264 & 0.057 & 0.21 & 0.63 & 0.00 & 1.00 & 0.7 & 4.9 & 325 & III\\
  & 0.6 & 0.90 & 0.81 & 0.11 & 0.194 & 0.215 & 0.022 & 0.10 & 0.50 & 0.00 & 1.00 & 0.5 & 5.2 & 100 & II\\
  & 0.6 & 0.80 & 0.69 & 0.17 & 0.131 & 0.160 & 0.027 & 0.17 & 0.43 & 0.00 & 1.00 & 0.5 & 5.8 & 195 & II\\
  & 0.6 & 0.70 & 0.56 &15.02 & 0.072 & 0.006 & 0.111 &17.09 & 0.33 & 0.28 & 0.37 & 5.3 & 6.6 & -- & II\\
A & 0.6 & 0.60 & 0.42 & 8.07 & 0.033 & 0.007 & 0.070 & 9.47 & 0.20 & 0.14 & 0.27 & 42.1& 5.1 & -- & I\\
  & 0.6 & 0.50 & 0.30 &14.70 & 0.014 & 0.002 & 0.038 &18.67 & 0.14 & 0.10 & 0.18 & 129.5&  6.0 & -- & I \\ \hline
  & 0.4 & 1.15& 1.10 & 0.25 & 0.505 & 0.000 & 0.000 & 0.27 & 1.00 & 1.00 & 1.00 & 1.0 &  --  & 112 & V \\
D & 0.4 & 1.00 & 0.92 & 1.02 & 0.299 & 0.188 & 0.197 & 1.05 & 0.61 & 0.14 & 0.95 & 1.5 & 4.9 & -- & II \\
  & 0.4 & 0.90 & 0.81 & 1.15 & 0.208 & 0.138 & 0.167 & 1.21 & 0.53 & 0.09 & 0.91 & 1.5 & 5.2 & -- & II \\
C & 0.4 & 0.70 & 0.57 & 7.00 & 0.070 & 0.013 & 0.111 & 8.26 & 0.31 & 0.22 & 0.39 & 7.6 & 6.6 & -- & II \\
  & 0.4 & 0.68& 0.55 &57.81 & 0.063 & 0.002 & 0.102 &51.80 & 0.30 & 0.28 & 0.32 & 8.8 & 6.9 & -- & II \\
  & 0.4 & 0.65& 0.51 & 8.13 & 0.052 & 0.008 & 0.089 &10.38 & 0.28 & 0.21 & 0.35 & 9.8 & 7.3 & -- & II \\
  & 0.4 & 0.60 & 0.45 & 4.93 & 0.037 & 0.010 & 0.071 & 6.54 & 0.23 & 0.12 & 0.31 &10.8 & 8.2 & -- & II \\
  & 0.4 & 0.50 & 0.32 &12.50 & 0.015 & 0.003 & 0.040 &13.98 & 0.13 & 0.09 & 0.19 &59.8 & 6.8 & -- & I \\ \hline
H & 0.2 & 1.12 & 1.07 & 4.23 & 0.438 & 0.040 & 0.175 & 4.30 & 0.93 & 0.86 & 0.98 & 1.0 & 4.8 & -- & IV\\
  & 0.2 & 1.00 & 0.94 & 2.68 & 0.289 & 0.093 & 0.257 & 2.76 & 0.59 & 0.36 & 0.80 & 2.4 & 4.9 & -- & II\\
  & 0.2 & 0.70 & 0.58 &11.00 & 0.071 & 0.009 & 0.113 &12.12 & 0.31 & 0.23 & 0.36 &10.4 & 6.8 & -- & II\\
  & 0.2 & 0.60 & 0.47 & 0.78 & 0.038 & 0.046 & 0.043 & 0.93 & 0.22 & 0.00 & 0.75 & 1.2 & 8.7 & -- & II\\
  & 0.2 & 0.50 & 0.34 & 8.65 & 0.016 & 0.003 & 0.041 &11.63 & 0.14 & 0.08 & 0.20 &83.7 & 12.8 & -- & I\\ \hline
  & 0.1 & 1.15 & 1.12 & 0.05 & 0.482 & 0.000 & 0.000 & 0.08 & 1.00 & 1.00 & 1.00 & 1.0 & -- & 188 & IV\\
  & 0.1 & 1.00 & 0.94 & 4.41 & 0.286 & 0.059 & 0.268 & 4.52 & 0.59 & 0.44 & 0.73 & 2.4 & 4.9 & -- & II\\
  & 0.1 & 0.70 & 0.59 & 9.55 & 0.071 & 0.010 & 0.113 &10.87 & 0.31 & 0.26 & 0.38 &15.1 & 7.0 & -- & II\\
  & 0.1 & 0.60 & 0.47 & 4.03 & 0.037 & 0.013 & 0.071 & 5.30 & 0.22 & 0.13 & 0.37 &20.3 & 8.8 & -- & II\\
B & 0.1 & 0.50 & 0.35 & 2.73 & 0.016 & 0.009 & 0.039 & 4.47 & 0.14 & 0.02 & 0.27 &69.6 & 12.1&  -- & I\\ \hline
G & 0.05& 1.15 & 1.12 & 13.90 & 0.478 & 0.006 & 0.086 & 14.03 & 0.99 & 0.97 & 0.99 & 1.0 &  4.8 & -- & IV \\
  & 0.05& 1.12 & 1.06 & 17.33 & 0.427 & 0.019 & 0.337 & 17.52 & 0.71 & 0.66 & 0.75 & 2.1 &  4.8 & --  & IV \\
  & 0.05& 1.00 & 0.94 & 6.82 & 0.284 & 0.038 & 0.272 & 7.00 & 0.58 & 0.48 & 0.67 & 5.5 & 4.9 & --  & II \\
  & 0.05& 0.70 & 0.59 &11.95 & 0.069 & 0.008 & 0.113 &14.39 & 0.29 & 0.24 & 0.35 &31.1 & 7.0 & --  & II \\
  & 0.05& 0.60 & 0.47 & 9.14 & 0.036 & 0.006 & 0.073 &11.40 & 0.22 & 0.17 & 0.29 &54.7 & 8.9 & --  & II \\
  & 0.05& 0.50 & 0.35 & 12.28 & 0.016 & 0.002 & 0.042 & 15.79 & 0.14 & 0.10 & 0.17 & 95.3 & 12.3 & -- & I
\enddata
\tablecomments{$\tau = 1.5 \times 10^7$ years, and $\alpha = 0.3$.  Periodic 
boundary conditions were used for all simulations.  The coupling coefficient 
and the gas surface density are given
in columns 1 and 2.  The statistics are based on the final $2.5 \times 10^8$ 
years of the $5 \times 10^9$ year simulations.  
Columns 3 and 5 give the spatio-temporal average
of the internal energy and the star formation rate (SFR).  Columns 6 and 7
give the standard deviation of the spatially averaged SFR and the temporal
average of the standard deviation of the SFR across the galaxy.  The ratio
of these two statistics $\chi_{\rm SFR}$ (see column 8) measures the degree
of global synchronization.  The equivalent statistic for the energy field
is given in column 4.  Columns 8, 9, 10 give the average, minimum and maximum
of the star formation filling factors (the fraction of lattice sites where
stars are forming on a given time step).  Column 12 is the number of disjoint
areas of star formation.  Column 13 is the average time between peaks (in
units of $10^7$ years) in the
global star formation rate.  Column 14 is the time required for global
synchronization to occur (i.e. $\chi_{\rm E} < 0.3$).  The phase as indicated
in figure 4 is given in column 15.}
\end{deluxetable}